\documentclass[preprint,aps]{revtex4}
\newcommand{\be}{\begin{equation}}
\newcommand{\ee}{\end{equation}}
\newcommand{\ba}{\begin{eqnarray}}
\newcommand{\ea}{\end{eqnarray}}

\begin{document}

\draft

\preprint{astro-ph/0309389}

\title{Charges and Magnetic Fields of the\\
       Slowly-rotating Neutron Stars}

\author{Hongsu Kim\footnote{e-mail : hongsu@astro.snu.ac.kr} and
Hyung Mok Lee\footnote{e-mail :hmlee@astro.snu.ac.kr}}

\address{Astronomy Program, SEES, Seoul National University, Seoul, 151-742, KOREA}

\author{Chul H. Lee\footnote{e-mail : chlee@hepth.hanyang.ac.kr} and
Hyun Kyu Lee\footnote{e-mail : hklee@hepth.hanyang.ac.kr}}

\address{Department of Physics, Hanyang University, Seoul, 133-791, KOREA}


\begin{abstract}
In association with the Goldreich-Julian's magnetic braking mechanism to explain the luminosity
of radio pulsars as a result of their spin-down, it is of some interest to explore how much of 
magnetic flux can actually penetrate the surface of the rotating neutron stars at least in idealized 
situations. In order to address this issue, one needs to figure out the amount of charge
on and the structure of magnetic field around a rotating neutron star. In the present
work, based on the solution-generating method given by Wald, the magnetic fields around
both the uncharged and (slightly) charged neutron star have been obtained. Particularly for the
charged neutron star, it has been demonstrated following again the argument by Wald that 
the neutron star will gradually accrete the charge until it reaches the equilibrium value 
$\tilde{Q}=2B_{0}J$.  Then next, the magnetic flux through one half of the
surface of the rotating neutron star has been computed as well. With the nonvanishing accretion 
charge having value in the range $0 < \tilde{Q} \leq 2B_{0}J$, the total magnetic flux through the 
neutron star has been shown to be greater than that without the accretion charge.
\end{abstract}


\maketitle

\narrowtext

\newpage
\begin{center}
{\rm\bf I. Introduction}
\end{center}

Radio pulsars are perhaps the oldest-known and the least energetic among all of the pulsar categories.
The thoretical study of these radio pulsars can be traced back to the 1969 work of Goldreich and 
Julian \cite{gj}. In their pioneering work, Goldreich and Julian argued that pulsars, which are
thought to be the rotating magnetized neutron stars, must have a {\it magnetosphere} with
charge-separated plasma. They then demonstrated that an electric force which is much
stronger than the gravitational force will be set up along the magnetic field and as a result, 
the surface charge layer cannot be in dynamical equilibrium. Then there appear steady current flows 
along the magnetic field lines which are taken to be uniformly rotating since they are firmly rooted 
in the crystalline crust of the pulsar surface. Although it was rather implicit in their original work,
this model for the pulsar electrodynamics in terms of the force-free magnetosphere does indeed
suggest that the luminosity of radio pulsars is due to the loss of their rotational energy (namely 
the spin-down) and its subsequent conversion into charged particle emission which eventually 
generates the radiation in the far zone. To elaborate on this last point, one should address the
issue of what exactly the working mechanism for the release of the extracted rotational
energy to infinitely far region would be. Indeed, if the field strength is large enough, say, of order
$B\sim 10^{12} (G)$ like the one observed near the radio pulsars, the vacuum surrounding the pulsar 
becomes unstable because any stray charged particles will be electrostatically
accelerated and will radiate and this radiation will in turn produce further charged particles in the 
form of electron-positron pairs. When charges are produced so freely, the electromagnetic field 
around the neutron star will become approximately force-free, namely a force-free
magnetosphere will be established. Therefore in reality, if the surrounding magnetic field is strong 
enough, charged particles are available due to the cascade production of electron-positron pairs and 
hence the extracted rotational energy of the neutron star may be released in the form of charged 
particle emission. Therefore in association with this Goldreich-Julian's mechanism for the spin-down
of rotating neutron stars to implicitly explain the pulsar luminosity, it is of some significance 
to explore how much of the magnetic field lines actually penetrate the surface of the neutron
stars at least in idealized situations. \\
Upon appreciating the idea of Goldreich and Julian for pulsar magnetosphere, the next natural
question would be ; what happens if we switch the central compact object from rotating neutron
stars to rotating (Kerr) black holes ?  Will there be force-free black hole magnetosphere instead ?
Indeed, such question has been asked and answered by Blandford and Znajek \cite{bz} long ago
which since then has been known as the Blandford-Znajek mechanism.
The way how some of the hole's rotational energy is carried off by the magnetic field lines 
threading the hole in the conventional Blandford-Znajek process can be summarized as follows. 
There the angular momentum and the rotational energy transport is
achieved in terms of the conservation of ``electromagnetic'' angular momentum and energy flux flowing 
from the hole to its nearby field particularly when the force-free condition is satisfied in the 
surrounding magnetosphere. It is also amusing to note that the Blandford-Znajek mechanism can be 
translated into a simple-minded circuit analysis. In this alternative picture, the current, flowing 
from near the pole toward the equator of the rotating black hole geometry and the
poloidal magnetic field together generate a ``magnetic braking'' torque which is antiparallel to the 
hole's spin direction. As a result, the rotating hole, which is a part of the circuit, spins down 
and its angular momentum and the rotational energy are extracted. 
For more details, we refer the reader to the original work of Blandford and Znajek \cite{bz} and 
some comprehensive review articles \cite{hklee} on this topic including the ``membrane paradigm'' 
approach \cite{membrane}. Therefore, both the Goldreich-Julian and the Blandford-Znajek mechanisms
can be dubbed ``magnetic braking'' mechanisms to explain the pulsar luminosity and the luminosity
of active galactic nuclei (AGNs) respectively as a result of their spin-downs.
As such, in order for both the Goldreich-Julian and the Blandford-Znajek processes
to work, the local magnetic field lines should be firmly rooted on the surface of the central compact
objects to carry away some of their rotational energies by means of this ``contact interaction''. \\
In our earlier work \cite{hongsu}, it has been explored how much of the magnetic flux can actually
penetrate the horizon of a rotating black hole in association with the Blandford-Znajek mechanism
for rotational energy extraction from a Kerr hole. And in doing so, there we made use of the algorithm
suggested by Wald to figure out the amount of charge on and the structure of magnetic field around
a rotating black hole. The results concerning these issues presented there for the case of a rotating
black hole can therefore be compared with those that will be presented here in this work for the case 
of a rotating neutron star since we shall in this work employ the algorithm of Wald's to determine
the amount of accretion charge on and the structure of magnetic field around the rotating neutron 
star as well. Thus it should be interesting to read the present work in 
parallel with our earlier work \cite{hongsu} on the Kerr black hole electrodynamics.

\begin{center}
{\rm\bf II. The Hartle-Thorne metric for the region exterior to the slowly-rotating
neutron stars}
\end{center}

We first begin by providing the rationale for treating the vicinity of a rotating neutron
star relativistically as a non-trivial curved spacetime. Since the pioneering proposals
by Gold \cite{gold} and by Pacini \cite{pacini}, it has by now been widely accepted that indeed pulsars
might be rotating magnetized neutron stars. Nevertheless, since then nearly all the studies 
on the pulsar electrodynamics have been performed by simply treating the region surrounding
the rotating neutron stars as being flat. This simplification may be sufficient just to gain
some insight into rough understanding of the origin of the pulsars' radiation. However, nearly
over thirty years have passed of the study of pulsar electrodynamics and it seems to be that
now we should look into the situation in a more careful and rigorous manner. In the present work,
we would like to take one step forward in this direction. We are particularly aimed at exploring
how much of the magnetic flux can actually penetrate the surface of a rotating neutron star and
to this end we shall first determine the amount of charge on and the structure of magnetic field 
around a slowly-rotating neutron star by treating its exterior region literally as a curved spacetime 
and employing the algorithm of Wald's as mentioned in the introduction. Indeed, thus far only a
handful of earlier studies concerning the electrodynamics of the rotating neutron stars appeared in 
the literature and they include the works by Muslimov and Tsygan, Muslimov and Harding 
\cite{muslimov} and by Prasanna and Gupta \cite{gupta}. Although their motivation, approach and
tools employed are completely different from those of our present work, Rezzolla et al. \cite{rezzolla}
also considered analytic solutions of Maxwell equations in the internal and external regions of a
slowly-rotating magnetized neutron star. \\
Indeed, we now
have a considerable amount of observed data for various species from radio pulsars \cite{gj} to
(anomalous) X-ray pulsars \cite{x-ray}. Even if we take the oldest-known radio pulsars for example,
it is not hard to realize that these objects are compact enough which really should be treated 
in general relativistic manner. To be more concrete, note that the values of the parameters
characterizing typical radio pulsars are ; $r_{0}({\rm radius}) \sim 10^{6} (cm)$, 
$M({\rm mass}) \sim 1.4 M_{\odot} \sim 2\times 10^{33} (g)$, 
$\tau({\rm pulsation ~period}) \sim 10^{-3} - 1 (sec)$, 
$B({\rm magnetic ~field ~strength}) \sim 10^{12} (G)$. 
Thus the Schwarzschild radius (gravitational radius) of a typical radio pulsar is estimated to
be $r_{Sch} = 2GM_{\odot}/c^2 \sim 3\times 10^{5} (cm)$ (where $G$ and $c$ denote the Newton's
constant and the speed of light respectively) and hence one ends up with the ratio
$r_{0}/r_{Sch} \sim 10^{6} (cm)/3\times 10^{5} (cm) \sim 3$. This simple argument indicates that indeed
even the radio pulsar (which is perhaps the least energetic among all of its species) is a
highly self-gravitating compact object that needs to be treated relativistically. Of course,
the ``relativistic'' treatment here means that the region surrounding the pulsar, namely the
pulsar magnetosphere has to be described by a curved spacetime rather than simply a flat
one. Then the question now boils down to ; what would be the relevant metric to describe
the vicinity of a rotating neutron star ?  Although it does not seem to be well-known, fortunately
we have such a metric of the region exterior to {\it slowly-rotating} relativistic stars such as
neutron stars, white dwarfs and supermassive stars and it is the one constructed long ago by
Hartle and Thorne \cite{ht}. Thus the Hartle-Thorne metric is a stationary axisymmetric solution to
the vacuum Einstein equation and in the present work, we shall take this Hartle-Thorne metric as 
a relavant one to represent the spacetime exterior to slowly-rotating neutron stars. One might then
wonder how the Hartle-Thorne metric for slowly-rotating neutron stars can be used to describe the
millisecond pulsars which are thought to be rapidly-rotating magnetized neutron stars. Here, the
``slowly-rotating'' really means that the relativistic stars like neutron stars are relatively
slowly rotating compared to the equal mass Kerr black hole which can rotate arbitrarily rapidly 
up to the maximal rotation $J = M^2$. Thus as we shall discuss later on in the concluding remarks, 
the angular speed of the rotating neutron star represented by the Hartle-Thorne metric is indeed
large enough to describe the millisecond pulsars. Now the Hartle-Thorne metric is given by
\begin{eqnarray}
ds^2 = -\alpha^2 dt^2 + g_{rr}dr^2 + g_{\theta\theta}d\theta^2 + g_{\phi\phi}
(d\phi + \beta^{\phi}dt)^2 
\end{eqnarray}
where
\begin{eqnarray}
\alpha^2 &=& \Delta R, ~~~\beta^{\phi} = -\omega = -{2J\over r^3}, \nonumber \\
g_{rr} &=& {S\over \Delta}, ~~~g_{\theta\theta} = r^2 A, 
~~~g_{\phi\phi} = \varpi^2 = r^2 A\sin^2 \theta, \\
g_{tt} &=& -[\alpha^2 - (\beta^{\phi})^2g_{\phi\phi}] = - [\Delta R - {4J^2\over r^4}A\sin^2 \theta],
\nonumber \\
g_{t\phi} &=& \beta^{\phi} g_{\phi\phi} = -{2J\over r}A\sin^2 \theta \nonumber
\end{eqnarray}
and
\begin{eqnarray}
\Delta &=& \left(1 - {2M\over r} + {2J^2\over r^4}\right), \nonumber \\
R &=& \left[1 + 2\left\{{J^2\over Mr^3}\left(1+{M\over r}\right) + {5\over 8}
{Q-J^2/M \over M^3}Q^2_{2}({r\over M}-1)\right\}P_{2}(\cos \theta)\right], \\
S &=& \left[1 - 2\left\{{J^2\over Mr^3}\left(1-{5M\over r}\right) + {5\over 8}
{Q-J^2/M \over M^3}Q^2_{2}({r\over M}-1)\right\}P_{2}(\cos \theta)\right], \nonumber \\
A &=& 1 + 2\left[-{J^2\over Mr^3}\left(1+{2M\over r}\right) \right. \nonumber \\
&&\left. + {5\over 8}{Q-J^2/M \over M^3}\left\{{2M\over [r^2(1-2M/r)]^{1/2}}Q^1_{2}({r\over M}-1)
- Q^2_{2}({r\over M}-1)\right\}\right]P_{2}(\cos \theta) \nonumber
\end{eqnarray}
with $M$, $J$ and $Q$ being the mass, the angular momentum and the
mass quadrupole moment of the (slowly) rotating neutron star respectively,
$P_{2}(\cos \theta) = (3\cos^2 \theta -1)/2$ being the Legendre polynomial, and 
$Q^{m}_{n}$ being the associated Legendre polynomial, namely,
\begin{eqnarray}
Q^{1}_{2}(z) &=& (z^2-1)^{1/2}\left[{3z^2 - 2\over z^2 - 1} 
- {3\over 2}z\log \left({z+1\over z-1}\right)\right], \\
Q^{2}_{2} (z) &=& \left[{3\over 2}(z^2 - 1)\log \left({z+1\over z-1}\right) 
- {3z^3 - 5z\over z^2 - 1}\right] \nonumber
\end{eqnarray}
and hence
\small
\begin{eqnarray}
Q^{1}_{2}({r\over M}-1) &=& {r\over M}\left(1-{2M\over r}\right)^{1/2}
\left[{3(r/M)^2(1-2M/r)+1 \over (r/M)^2(1-2M/r)}
+ {3\over 2}{r\over M}\left(1-{M\over r}\right)\log \left(1-{2M\over r}\right)\right], \\
Q^{2}_{2}({r\over M}-1) &=& -\left[{3\over 2}\left({r\over M}\right)^2\left(1-{2M\over r}\right)
\log \left(1-{2M\over r}\right) + 
{{M/r}(1-M/r)\left\{3(r/M)^2(1-2M/r)-2\right\} \over 
(1-2M/r)}\right]. \nonumber
\end{eqnarray}
\normalsize
As is well-known, the only known exact metric solution exterior to a rotating object is the
Kerr metric \cite{kerr}. Thus it would be worth clarifying the relation of the Hartle-Thorne metric for
slowly-rotating relativistic stars given above to the Kerr metric. As Hartle and Thorne \cite{ht} pointed
out, take the Kerr metric given in Boyer-Lindquist coordinate and expand it to second order
in angular velocity (namely, the angular shift $\beta^{\phi}$) followed by a coordinate
transformation in the $(r, \theta)$-sector,
\begin{eqnarray}
&&r \to  r\left[1-{a^2\over 2r^2}\left\{(1+{2M\over r})(1-{M\over r})+\cos^2 \theta
(1-{2M\over r})(1+{3M\over r})\right\}\right], \nonumber \\
&&\theta \to  \theta - a^2 \cos \theta \sin \theta {1\over 2r^2}(1+{2M\over r}) 
\end{eqnarray}
where $a=J/M$.
Then one can realize that the resulting expanded Kerr metric coincides with the particular case
$Q=J^2/M$ (with $Q$ being the mass quadrupole moment of the rotating object) of the Hartle-Thorne
metric. Therefore, in general this Hartle-Thorne metric is {\it not} a slow-rotation limit of Kerr 
metric. Rather, the slow-rotation limit of Kerr metric is a special case of this more general 
Hartle-Thorne metric. As a result, the Hartle-Thorne metric with an arbitrary value of the mass
quadrupole moment $Q$ can generally describe a (slowly-rotating) neutron star of any shape
(as long as it retains the axisymmetry). \\
Now we turn to the choice of an orthonormal tetrad frame. And we shall particularly choose the
Zero-Angular-Momentum-Observer (ZAMO) \cite{zamo} frame which is a sort of fiducial observer (FIDO) frame.
Generally speaking, in order to represent a given background geometry, one needs to first 
choose a coordinate system in which the metric is to be given and next, in order to obtain 
physical components of a tensor (such as the electric and magnetic field values), one has to 
select a tetrad frame (in a given coordinate system) to which the tensor components are to be
projected. As is well-known, the orthonormal tetrad is a set of four mutually orthogonal unit 
vectors at each point in a given spacetime which give the directions of the four axes of 
locally-Minkowskian coordinate system. Such an orthonormal tetrad associated with the 
Hartle-Thorne metric given above may be chosen as  
$e^{A}= e^{A}_{\mu}dx^{\mu} = (e^{0}, e^{1}, e^{2}, e^{3})$,
\ba
e^{0} &=& \alpha dt = (\Delta R)^{1/2}dt, \nonumber  \\
e^{1} &=& g^{1/2}_{rr}dr = \left({S\over \Delta}\right)^{1/2}dr, \\
e^{2} &=& g^{1/2}_{\theta \theta}d\theta = rA^{1/2}d\theta, \nonumber  \\
e^{3} &=& g^{1/2}_{\phi \phi}(d\phi + \beta^{\phi}dt) =
r\sin \theta A^{1/2}\left[d\phi - {2J\over r^3}dt\right] \nonumber
\ea
and its dual basis is given by
$e_{A}= e^{\mu}_{A}\partial_{\mu} = 
(e_{0}=e_{(t)}, e_{1}=e_{(r)}, e_{2}=e_{(\theta)}, e_{3}=e_{(\phi)})$,
\ba
e_{0} &=& {1\over \alpha}(\partial_{t}-\beta^{\phi} \partial_{\phi}) = 
(\Delta R)^{-1/2}\left[\partial_{t}+\frac{2J}{r^3}\partial_{\phi}\right], \nonumber  \\
e_{1} &=& g^{-1/2}_{rr}\partial_{r} = \left(\frac{\Delta}{S}\right)^{1/2}\partial_{r}, \\
e_{2} &=& g^{-1/2}_{\theta \theta}\partial_{\theta} = 
\frac{1}{rA^{1/2}}\partial_{\theta}, \nonumber  \\
e_{3} &=& g^{-1/2}_{\phi \phi}\partial_{\phi} = 
\frac{1}{rA^{1/2}\sin \theta}\partial_{\phi}. \nonumber
\ea
The local, stationary observer at rest in this orthonormal tetrad frame $e^{A}$ has the worldline
given by $\{dr=0, ~d\theta=0, ~(d\phi + \beta^{\phi}dt)=0\}$ which is orthogonal to spacelike 
hypersurfaces and has orbital angular velocity given by
\be
\omega = \frac{d\phi}{dt} = -\beta^{\phi} = -\frac{g_{t\phi}}{g_{\phi \phi}}
= \frac{2J}{r^3}.
\ee
This is the long-known {\it Lense-Thirring} precession \cite{lt} angular velocity arising due to
the ``dragging of inertial frame'' effect of a stationary axisymmetric spacetime. Indeed, it is
straightforward to demonstrate that this orthonormal tetrad observer can be identified with a
ZAMO carrying zero intrinsic angular momentum with it. 
To this end, recall that when a spacetime metric
possesses a rotational (azimuthal) isometry, there exists associated rotational Killing field
$\psi^{\mu} = (\partial /\partial \phi)^{\mu} = \delta^{\mu}_{\phi}$ such that the inner product
of it with the tangent (velocity) vector $u^{\mu} = dx^{\mu}/d\tau$ (with $\tau $ denoting the
particle's proper time) of the geodesic of a test particle is constant along the geodesic, i.e.,
\ba
\tilde{L} &=& g_{\alpha \beta}\psi^{\alpha}u^{\beta} = 
g_{\phi t}\psi^{\phi}u^{t} + g_{\phi \phi}\psi^{\phi}u^{\phi} \nonumber \\
&=& g_{\phi t}\frac{dt}{d\tau} + g_{\phi \phi}\frac{d\phi}{d\tau}.
\ea
Now, particularly when the local, stationary observer, which here is taken to be a test particle, 
carries zero angular momentum, $\tilde{L} = 0$, its angular velocity becomes
\be
\omega = \frac{d\phi}{dt} = \frac{(d\phi/d\tau)}{(dt/d\tau)} = 
-\frac{g_{t\phi}}{g_{\phi \phi}} = -\beta^{\phi}
\ee
and this confirms the identification of the local observer at rest in this orthonormal tetrad
frame given above with a ZAMO.

\begin{center}
{\rm\bf III. Solution Generating Method by Wald}
\end{center}

{\bf 1. Wald field}

From the general properties of Killing fields \cite{papa} ; {\it a Killing vector in
a vacuum spacetime generates a solution of Maxwell's equations in the
background of that vacuum spacetime}, long ago, Wald \cite{wald} constructed a stationary
axisymmetric solution of Maxwell's equations in Kerr black hole spacetime. 
To be a little more concrete, Wald's construction is based on the following two
statements : \\
(A) {\it The axial Killing vector $\psi^{\mu} = (\partial/\partial \phi)^{\mu}$
generates a stationary axisymmetric test electromagnetic field which asymptotically
approaches a uniform magnetic field, has no magnetic monopole moment and has 
charge $= 4J$,} \\
$F_{\psi} = d\psi $ \\
where ``$d$'' denotes the exterior derivative and $J$ is the angular momentum of a 
Kerr black hole. \\
(B) {\it The time translational Killing vector $\xi^{\mu} = (\partial/\partial t)^{\mu}$
generates a stationary axisymmetric test electromagnetic field which vanishes
asymptotically, has no magnetic monopole moment and has
charge $= -2M$,} \\
$F_{\xi} = d\xi $ \\
where $M$ is the mass of the Kerr hole. \\
Apparently, this solution generating method of Wald's can be applied in a straightforward manner
to the construction of a stationary axisymmetric solution of Maxwell's equations in a spacetime
surrounding a slowly-rotating neutron star represented by the Hartle-Thorne metric given earlier.
Thus we now look for the solution of the electromagnetic test field $F$ which occurs when 
a stationary axisymmetric slowly-rotating neutron star of mass $M$ and angular momentum $J$
(represented by the Hartle-Thorne metric) 
is placed in an originally uniform magnetic field of strength $B_{0}$ aligned along the symmetry 
axis of the neutron star. And for now we consider the case when the electric charge is absent. 
Obviously, then the solution can be readily written down by referring to the statements (A) and (B) 
above as follows 
\begin{eqnarray}
F = {1\over 2}B_{0} \left[F_{\psi}-\left({4J\over -2M}\right)F_{\xi}\right]
= {1\over 2}B_{0} \left[d\psi + {2J\over M} d\xi \right].
\end{eqnarray}
Then using ; $F={1\over 2}F_{\mu\nu}dx^{\mu}\wedge dx^{\nu}$, 
$\psi = \psi_{\mu}dx^{\mu}$, $\xi = \xi_{\nu}dx^{\nu}$, with
\begin{eqnarray}
\xi_{\mu} &=& g_{\mu\nu}\xi^{\nu}=g_{\mu\nu}\delta^{\nu}_{t}=g_{\mu t},\\
\psi_{\mu} &=& g_{\mu\nu}\psi^{\nu}=g_{\mu\nu}\delta^{\nu}_{\phi}=g_{\mu\phi}
\nonumber
\end{eqnarray}
and for the Hartle-Thorne metric given in the previous subsection, the solution
above can be written in a concrete form as 
\begin{eqnarray}
F &=& \frac{1}{2}B_{0}\left[(\partial_{r}g_{t\phi})+\frac{2J}{M}(\partial_{r}g_{tt})\right]
(dr \wedge dt) 
  + \frac{1}{2}B_{0}\left[(\partial_{\theta}g_{t\phi})+\frac{2J}{M}(\partial_{\theta}g_{tt})\right]
(d\theta \wedge dt) \nonumber \\
  &+& \frac{1}{2}B_{0}\left[(\partial_{r}g_{\phi \phi})+\frac{2J}{M}(\partial_{r}g_{\phi t})\right]
(dr \wedge d\phi) 
  + \frac{1}{2}B_{0}\left[(\partial_{\theta}g_{\phi \phi})+\frac{2J}{M}(\partial_{\theta}g_{\phi t})\right]
(d\theta \wedge d\phi)] 
\end{eqnarray}
and here
\small
\begin{eqnarray}
(\partial_{r}g_{tt}) &=& 
- \left(\frac{2M}{r^2}-\frac{8J^2}{r^5}\right)R + 2\Delta \left[\frac{J^2}{Mr^4}
\left(3+\frac{4M}{r}\right) - \frac{5}{8} \frac{Q-J^2/M}{M^3}Q^{'2}_{2}(\frac{r}{M}-1)\right]P_{2}(\cos \theta) \nonumber \\
&-& \frac{16J^2}{r^5}A\sin^2 \theta
+ \frac{8J^2}{r^4}\sin^2 \theta \left[\frac{J^2}{Mr^4}\left(3+\frac{8M}{r}\right) + \frac{5}{8} \frac{Q-J^2/M}{M^3}\times \right. \\
&&\left. \left\{\frac{-2M(1-M/r)}{r^2 (1-2M/r)^{3/2}}Q^{1}_{2}(\frac{r}{M}-1)
+ \frac{2M}{[r^2 (1-2M/r)]^{1/2}}Q^{'1}_{2}(\frac{r}{M}-1) - Q^{'2}_{2}(\frac{r}{M}-1)\right\}\right]P_{2}(\cos \theta), \nonumber \\
(\partial_{\theta}g_{tt}) &=& 
6\Delta \left[\frac{J^2}{Mr^3}
\left(1+\frac{M}{r}\right) + \frac{5}{8} \frac{Q-J^2/M}{M^3}Q^{2}_{2}(\frac{r}{M}-1)\right]\sin \theta \cos \theta \nonumber \\
&+& \frac{8J^2}{r^4}A\sin \theta \cos \theta
- \frac{24J^2}{r^4}\left[-\frac{J^2}{Mr^3}\left(1+\frac{2M}{r}\right)  \right. \\
&&\left. + \frac{5}{8} \frac{Q-J^2/M}{M^3}\left\{
+ \frac{2M}{[r^2 (1-2M/r)]^{1/2}}Q^{1}_{2}(\frac{r}{M}-1) - Q^{2}_{2}(\frac{r}{M}-1)\right\}\right]\sin^3 \theta \cos \theta, \nonumber \\
(\partial_{r}g_{t\phi}) &=& 
\frac{2J}{r^2}A\sin^2 \theta
- \frac{4J}{r}\sin^2 \theta \left[\frac{J^2}{Mr^4}\left(3+\frac{8M}{r}\right) + \frac{5}{8} \frac{Q-J^2/M}{M^3}\times \right. \\
&&\left. \left\{\frac{-2M(1-M/r)}{r^2 (1-2M/r)^{3/2}}Q^{1}_{2}(\frac{r}{M}-1)
+ \frac{2M}{[r^2 (1-2M/r)]^{1/2}}Q^{'1}_{2}(\frac{r}{M}-1) - Q^{'2}_{2}(\frac{r}{M}-1)\right\}\right]P_{2}(\cos \theta), \nonumber \\
(\partial_{\theta}g_{t\phi}) &=& 
- \frac{4J}{r}A\sin \theta \cos \theta
+ \frac{12J}{r}\left[-\frac{J^2}{Mr^3}\left(1+\frac{2M}{r}\right) \right. \\
&&\left. + \frac{5}{8} \frac{Q-J^2/M}{M^3}\left\{
+ \frac{2M}{[r^2 (1-2M/r)]^{1/2}}Q^{1}_{2}(\frac{r}{M}-1) - Q^{2}_{2}(\frac{r}{M}-1)\right\}\right]\sin^3 \theta \cos \theta, \nonumber \\
(\partial_{r}g_{\phi \phi}) &=& 
2rA\sin^2 \theta
+ 2r^2 \sin^2 \theta \left[\frac{J^2}{Mr^4}\left(3+\frac{8M}{r}\right) + \frac{5}{8} \frac{Q-J^2/M}{M^3}\times \right. \\
&&\left. \left\{\frac{-2M(1-M/r)}{r^2 (1-2M/r)^{3/2}}Q^{1}_{2}(\frac{r}{M}-1)
+ \frac{2M}{[r^2 (1-2M/r)]^{1/2}}Q^{'1}_{2}(\frac{r}{M}-1) - Q^{'2}_{2}(\frac{r}{M}-1)\right\}\right]P_{2}(\cos \theta), \nonumber \\
(\partial_{\theta}g_{\phi \phi}) &=& 
2r^2 A\sin \theta \cos \theta
- 6r^2 \left[-\frac{J^2}{Mr^3}\left(1+\frac{2M}{r}\right) \right. \\
&&\left. + \frac{5}{8} \frac{Q-J^2/M}{M^3}\left\{
+ \frac{2M}{[r^2 (1-2M/r)]^{1/2}}Q^{1}_{2}(\frac{r}{M}-1) - Q^{2}_{2}(\frac{r}{M}-1)\right\}\right]\sin^3 \theta \cos \theta \nonumber 
\end{eqnarray}
\normalsize
where
\small
\begin{eqnarray}
&&Q^{'1}_{2}(\frac{r}{M}-1) = \frac{d}{dr}Q^{1}_{2}(\frac{r}{M}-1)  \\
&&={1\over r}\left(1-{2M\over r}\right)^{-1/2}
\left[{{M/r}(1-M/r)\left\{6(r/M)^2(1-2M/r)-1\right\} \over (1-2M/r)} \right. \nonumber \\
&&\left. + {3\over 2}\left\{2\left({r\over M}\right)^2\left(1-{2M\over r}\right)+1\right\}\log \left(1-{2M\over r}\right)\right], \nonumber \\
&&Q^{'2}_{2}({r\over M}-1) = \frac{d}{dr}Q^{2}_{2}(\frac{r}{M}-1) \\
&&= -\frac{1}{M}\left[6 + 3\left({r\over M}\right)\left(1-{M\over r}\right)
\log \left(1-{2M\over r}\right) + 
2\left({M\over r}\right)^2(1-{2M\over r})^{-1} + 4\left({M\over r}\right)^4(1-{2M\over r})^{-2}\right]. \nonumber
\end{eqnarray}
\normalsize
Thus this can be thought of as the Wald-type field for the case of a rotating neutron star.
Of course, a caution needs to be exercised and it is the point that this Maxwell field is correct
only for a slowly-rotating neutron star but not for a neutron star with arbitrary rotation.
\\
{\bf 2. Wald charge}
\\
We now turn to the issue of {\it charge accretion} onto the (slowly-rotating) neutron star immersed in a 
magnetic field surrounded by an ionized interstellar medium (``plasma''). We shall essentially
follow the argument again given by Wald \cite{wald} and to do so, we first need to know the physical components
of electric and magnetic fields. This can be achieved by projecting the Maxwell field tensor
given above in eq.(14) onto the ZAMO tetrad frame. Thus using the dual to the ZAMO tetrad,
$e_{A}=(e_{0}=e_{(t)}, e_{1}=e_{(r)}, e_{2}=e_{(\theta)}, e_{3}=e_{(\phi)})$, 
given earlier in eq.(8), we can now read off the ZAMO tetrad components of $F_{\mu \nu}$ as
\ba
F_{10} &=& F_{\mu\nu} e^{\mu}_{1}e^{\nu}_{0} \nonumber \\
      &=& \left(\frac{\Delta}{S}\right)^{1/2}\left[(\Delta R)^{-1/2}\frac{1}{2}B_{0}
\left\{(\partial_{r}g_{t\phi})+\frac{2J}{M}(\partial_{r}g_{tt})\right\} 
 +(\Delta R)^{-1/2}\frac{2J}{r^3}\frac{1}{2}B_{0}
\left\{(\partial_{r}g_{\phi \phi})+\frac{2J}{M}(\partial_{r}g_{\phi t})\right\}\right], \nonumber \\ 
F_{20} &=& F_{\mu\nu} e^{\mu}_{2}e^{\nu}_{0} \nonumber \\
      &=&  \frac{1}{rA^{1/2}}\left[(\Delta R)^{-1/2}\frac{1}{2}B_{0}
\left\{(\partial_{\theta}g_{t\phi})+\frac{2J}{M}(\partial_{\theta}g_{tt})\right\} 
 +(\Delta R)^{-1/2}\frac{2J}{r^3}\frac{1}{2}B_{0}
\left\{(\partial_{\theta}g_{\phi \phi})+\frac{2J}{M}(\partial_{\theta}g_{\phi t})\right\}\right], \nonumber \\
F_{30} &=& F_{\mu\nu} e^{\mu}_{3}e^{\nu}_{0} = 0, \\
F_{12} &=& F_{\mu\nu} e^{\mu}_{1}e^{\nu}_{2} = 0, \nonumber \\
F_{13} &=& F_{\mu\nu} e^{\mu}_{1}e^{\nu}_{3} \nonumber \\
      &=&  \left(\frac{\Delta}{S}\right)^{1/2} \frac{1}{rA^{1/2}\sin \theta}
\frac{1}{2}B_{0}\left\{(\partial_{r}g_{\phi \phi})+\frac{2J}{M}(\partial_{r}g_{\phi t})\right\}, \nonumber \\
F_{23} &=& F_{\mu\nu} e^{\mu}_{2}e^{\nu}_{3} \nonumber \\
      &=& \frac{1}{rA^{1/2}}\frac{1}{rA^{1/2}\sin \theta}
\frac{1}{2}B_{0}\left\{(\partial_{\theta}g_{\phi \phi})+\frac{2J}{M}(\partial_{\theta}g_{\phi t})\right\}
            \nonumber                          
\ea
where, $(\partial_{r}g_{t\phi})$, $(\partial_{r}g_{tt})$, $(\partial_{\theta}g_{t\phi})$,
$(\partial_{\theta}g_{tt})$, $(\partial_{r}g_{\phi \phi})$ and
$(\partial_{\theta}g_{\phi \phi})$ are as given above in eqs. (15)-(20).  \\
Here, consider particularly the radial component of the electric field (as observed by a local
observer in this ZAMO tetrad frame) which, along the symmetry axis ($\theta = 0, ~\pi$) of the 
slowly-rotating neutron star, becomes 
\begin{eqnarray}
&&E_{\hat{r}} = E_{1} = F_{10} \\
&&= \frac{1}{2}B_{0}(RS)^{-1/2}\frac{2J}{M}\left[-\left(\frac{2M}{r^2}-\frac{8J^2}{r^5}\right)R
+ 2\Delta \left\{\frac{J^2}{Mr^4}\left(3+\frac{4M}{r}\right) - \frac{5}{8}\frac{Q-J^2/M}{M^3}
Q^{'2}_{2}(\frac{r}{M}-1)\right\}\right]. \nonumber
\end{eqnarray}
Note that from this expression for the radial component of the electric field, it is {\it not}
possible to determine the sign of charge accreted in terms of the relative direction between the angular
momentum of the neutron star and the external magnetic field. And this point is in contrast to
what happens in the case of Kerr black hole\cite{wald, hongsu} studied previously.
To be a little more concrete, there it has been realized that the radial component of the
electric field for the case of Kerr black hole is radially {\it inward/outward}
if the hole's axis of rotation and the external magnetic field are {\it parallel/antiparallel}.
Put differently, this implies that if the spin of the hole and the magnetic field are {\it parallel},
then {\it positively charged} particles on the symmetry axis of the hole will be
pulled into the hole while if the spin of the hole and the magnetic field are {\it antiparallel},
{\it negatively charged} particles on the symmetry axis of the hole will be
pulled into the hole. In this manner a rotating black hole will ``selectively'' accrete charged
particles until it builds up ``equilibrium'' net charge. For the present case of rotating neutron
star, however, an argument of this sort cannot be presented due essentially to the added complexity 
of the Hartle-Thorne metric compared to the Kerr metric. Nevertheless, one can still ask and answer
the next natural question ; how then can the equilibrium net charge be determined ?  
And to answer this question, we resort to the ``injection energy'' argument originally proposed by 
Carter \cite{carter} for the Kerr black hole case. Again, it can be applied equally well to the present case
of (slowly-rotating) neutron star. Recall first that the energy of a charged 
particle in a stationary spacetime with the time translational isometry generated by the Killing 
field $\xi^{\mu}=(\partial /\partial t)^{\mu}$ in the presence of a stationary electromagnetic 
field is given by 
\begin{eqnarray}
\varepsilon = - p_{\alpha}\xi^{\alpha} = - g_{\alpha\beta}p^{\alpha}\xi^{\beta} \nonumber
\end{eqnarray}
with $p^{\mu}={\tilde m}u^{\mu}-eA^{\mu}$ being the 4-momentum of the charged particle with mass
and charge $\tilde{m}$ and $e$ respectively. Now if we lower the charged particle down the
symmetry axis into the rotating neutron star, the change in electrostatic energy of the particle will be
\begin{eqnarray}
\delta \varepsilon = \varepsilon_{final} - \varepsilon_{initial} =
eA_{\alpha}\xi^{\alpha}|_{horizon} - eA_{\alpha}\xi^{\alpha}|_{\infty}.
\end{eqnarray}
Now, if $\delta \varepsilon < 0 \rightarrow $ it will be energetically favorable for the neutron star
to accrete particles with this charge whereas if $\delta \varepsilon > 0 \rightarrow $  
it will accrete particles with opposite charge. In either case, the rotating neutron star will
selectively accrete charges until $A^{\mu}$ is changed sufficiently that the electrostatic ``injection
energy'' $\delta \varepsilon$ is reduced to zero. We are then ready to determine, by this injection
energy argument due to Carter, what the equilibrium net charge accreted onto the neutron star would be.
In the discussion of Wald-type field given above, we only restricted ourselves to the case of solutions to
Maxwell equation in the background of {\it uncharged} stationary axisymmetric neutron star spacetime
and it was given by eq.(12). Now we need the solution when the stationary axisymmetric neutron star is
slightly charged via charge accretion process described above. Then according to the {\it statement
(B)} in the discussion of Wald field given earlier, there can be at most one more perturbation of a
stationary axisymmetric vacuum (neutron star) spacetime which corresponds to adding an electric 
charge $\tilde{Q}$ to the neutron star and it is nothing but to linearly superpose the solution 
$(-\tilde{Q}/2M)F_{\xi} = (-\tilde{Q}/2M)d\xi$ to the solution given in eq.(12) to get 
\begin{eqnarray}
F = \frac{1}{2}B_{0}[d\psi + \frac{2J}{M}d\xi]-\frac{\tilde{Q}}{2M}d\xi,
\end{eqnarray} 
which, in terms of the gauge potental, amounts to
\begin{eqnarray}
A_{\mu} = {1\over 2}B_{0}(\psi_{\mu}+{2J\over M}\xi_{\mu})-{\tilde{Q}\over 2M}\xi_{\mu}.
\end{eqnarray}
Then the electrostatic injection energy can be computed as
\begin{eqnarray}
\delta \varepsilon &=& eA_{\alpha}\xi^{\alpha}|_{horizon} - eA_{\alpha}\xi^{\alpha}|_{\infty}
\nonumber \\
&=& e\left(\frac{B_{0}J}{M}-\frac{\tilde{Q}}{2M}\right)[1 - \Delta R|_{r=r_{0}, \theta =0, \pi}]
\end{eqnarray}
where we used $\xi_{\mu}\xi^{\mu} = -\Delta R|_{r=r_{0}, \theta =0, \pi}$ 
(at $r=r_{0}, \theta =0, \pi $),  $\xi_{\mu}\xi^{\mu} = -1$ (at $r\to \infty, \theta =0, \pi $)
and $\psi_{\mu}\xi^{\mu} = 0$ (on the symmetry axis $\theta =0, \pi $) and $r_{0}$ denotes the 
radius of the neutron star along the symmetry axis.
Thus one may conclude that a rotating neutron star in a uniform magnetic field will accrete charge until
the gauge potential evolves to a value at which $\delta \varepsilon = 0$ yielding the equilibrium
net charge as $\tilde{Q} = 2B_{0}J$. Thus this amount of charge may be called Wald-type charge and it is
particularly interesting to note that it turns out to be the same as the ``Wald charge'' (the 
equilibrium net charge) for the case of Kerr black hole \cite{wald, hongsu}. And this result appears
to imply that the amount of equilibrium net charge $\tilde{Q} = 2B_{0}J$ might be a generic value once a
general relativistic rotating object gets charged via accretion. There, however, does exist a
distinction between the case of Kerr black hole and that of a rotating neutron star.
That is, from eq.(28), the rotating neutron star will eventually accrete the equilibrium net charge
of $\tilde{Q} = 2B_{0}J$ only if $[1 - \Delta R|_{r=r_{0}, \theta =0, \pi}] \neq 0$. Certainly, this feature
was absent in the case of Kerr black hole and can be attributed to the different structure of
Hartle-Thorne metric describing the region surrounding the rotating neutron star.
Therefore, particularly if the values of mass ($M$), angular momentum ($J$), mass quadrupole 
moment ($Q$) and radius along the symmetry axis ($r_{0}$) are such that they satisfy
\begin{eqnarray}
&&\Delta R|_{r=r_{0}, \theta =0, \pi} \\
&&= \left(1 - {2M\over r_{0}} + {2J^2\over r^4_{0}}\right)
\left[1 + 2\left\{{J^2\over Mr^3_{0}}\left(1+{M\over r_{0}}\right) + {5\over 8}
{Q-J^2/M \over M^3}Q^2_{2}({r_{0}\over M}-1)\right\}\right] = 1 \nonumber
\end{eqnarray}
then it is ``energetically unlikely'' that a slowly-rotating neutron star would accrete
any charge and we find it a peculiar feature which has no parallel in the Kerr black hole case. 
It, however, should be taken with some caution (i.e., should not be taken as being conclusive) since
the injection energy argument due to Carter that leads to this conclusion is, rigorously speaking,
indeed restricted to the situation along the symmetry axis $\theta = 0, \pi$ and at this point it
is not so obvious that the same would be true off the symmetry axis as well. \\
Next, recall that we announced from the beginning that we shall consider the case when the charge
accreted on the neutron star is small enough not to distort the background Hartle-Thorne geometry. Now
we provide the rationale that this is indeed what can actually happen. To do so, we first
assume that just like in the case of Kerr black hole (a particular slow-rotation limit of which is
the Hartle-Thorne geometry), in the present Hartle-Thorne spacetime, $J\leq M^2$. 
Then using the fact that the typical value of the charge accreted on
the rotating neutron star is $\tilde{Q}=2B_{0}J$ as just described, its charge-to-mass
ratio has an upper bound
\begin{eqnarray}
{\tilde{Q} \over M} = 2B_{0}\left({J\over M}\right) \leq 2B_{0}M = 2\left({B_{0}\over 10^{15}(G)}\right)
\left({M\over M_{\odot}}\right)10^{-5} \nonumber
\end{eqnarray}
where in the last equality we have converted $B_{0}$ and $M$ from geometrized units to
solar-mass units and gauss \cite{hongsu}. Thus for the typical case of radio pulsar ; with mass 
$M\sim ~M_{\odot}$ in the surrounding magnetic field of strength $B_{0}\sim 10^{12} (G)$, 
$\tilde{Q}/M \sim 10^{-8} << 1$.  Thus as we can see in this radio pulsar case,
the charge-to-mass ratio of the associated rotating neutron star is small enough not to disturb 
the geometry itself. Thus we can safely employ the solution-generating method suggested by Wald 
to construct the solution to the Maxwell equations when some amounts of charges are around 
to which we now turn.  

\begin{center}
{\rm\bf IV. Stationary axisymmetric Maxwell field around a ``slightly charged'' and
slowly-rotating neutron star}
\end{center}

As discussed in the previous subsection, the stationary
axisymmetric solution to the Maxwell equation in the background of a rotating neutron star with charge $\tilde{Q}$
accreted in a originally uniform magnetic field can be constructed as    
\ba
F =\frac{1}{2}B_{0}[F_{\psi}-(\frac{4J}{-2M})F_{\xi}]+(\frac{\tilde{Q}}{-2M})F_{\xi}
=\frac{1}{2}B_{0}[d\psi + \frac{2J}{M}(1-\frac{\tilde{Q}}{2B_{0}J})d\xi]. 
\ea
Again for the Hartle-Thorne metric for a slowly-rotating neutron star given in the previous 
subsection, the solution above can be written as 
\ba
F &=& \frac{1}{2}B_{0}\left[(\partial_{r}g_{t\phi})+
\frac{2J}{M}\left(1 - \frac{\tilde{Q}}{2B_{0}J}\right)(\partial_{r}g_{tt})\right]
(dr \wedge dt) \nonumber \\
  &+& \frac{1}{2}B_{0}\left[(\partial_{\theta}g_{t\phi})+
\frac{2J}{M}\left(1 - \frac{\tilde{Q}}{2B_{0}J}\right)(\partial_{\theta}g_{tt})\right]
(d\theta \wedge dt) \\
  &+& \frac{1}{2}B_{0}\left[(\partial_{r}g_{\phi \phi})+
\frac{2J}{M}\left(1 - \frac{\tilde{Q}}{2B_{0}J}\right)(\partial_{r}g_{\phi t})\right]
(dr \wedge d\phi) \nonumber \\
  &+& \frac{1}{2}B_{0}\left[(\partial_{\theta}g_{\phi \phi})+
\frac{2J}{M}\left(1 - \frac{\tilde{Q}}{2B_{0}J}\right)(\partial_{\theta}g_{\phi t})\right]
(d\theta \wedge d\phi)] \nonumber
\ea
and here, $(\partial_{r}g_{t\phi})$, $(\partial_{r}g_{tt})$, $(\partial_{\theta}g_{t\phi})$,
$(\partial_{\theta}g_{tt})$, $(\partial_{r}g_{\phi \phi})$ and
$(\partial_{\theta}g_{\phi \phi})$ are as given earlier in the previous section.  \\
Obviously, in order to have some insight into the nature of this solution to the Maxwell equations,
one may wish to obtain physical components of electric field and magnetic induction. And this can 
only be achieved by projecting the Maxwell field tensor above onto an appropriate tetrad frame
like ZAMO tetrad which, for the Hartle-Thorne metric being considered, can be well-defined as 
mentioned earlier. The ZAMO tetrad is well-known and widely employed in 
various anaysis in the literature. ZAMO is a fiducial observer following a timelike geodesic 
orthogonal to spacelike hypersurfaces. And this implies that its 4-velocity is just the timelike
ZAMO tetrad $u^{\mu} = e^{\mu}_{0}$ given earlier in eq.(8). Thus upon projecting the Maxwell field 
tensor components on
the ZAMO tetrad frame, the physical components of electric and magnetic fields can be read off
as $E_{i}=F_{i0}=F_{\mu\nu}(e^{\mu}_{i}e^{\nu}_{0})$ and $B_{i}=\epsilon_{ijk}F^{jk}/2=
\epsilon_{ijk}F^{\mu\nu}(e^{\mu}_{j}e^{\nu}_{k})/2$, respectively. Namely,   
from the dual to the ZAMO tetrad, $e_{A}=(e_{0}=e_{(t)}, e_{1}=e_{(r)}, 
e_{2}=e_{(\theta)}, e_{3}=e_{(\phi)})$, given earlier in eq.(8),
\ba
F_{10} &=& F_{\mu\nu} e^{\mu}_{1}e^{\nu}_{0} \nonumber \\
      &=& \left(\frac{\Delta}{S}\right)^{1/2}\left[(\Delta R)^{-1/2}\frac{1}{2}B_{0}
\left\{(\partial_{r}g_{t\phi})+\frac{2J}{M}\left(1 - \frac{\tilde{Q}}{2B_{0}J}\right)
(\partial_{r}g_{tt})\right\} \right.  \nonumber \\
&&\left. +(\Delta R)^{-1/2}\frac{2J}{r^3}\frac{1}{2}B_{0}
\left\{(\partial_{r}g_{\phi \phi})+\frac{2J}{M}\left(1 - \frac{\tilde{Q}}{2B_{0}J}\right)
(\partial_{r}g_{\phi t})\right\}\right], \nonumber \\ 
F_{20} &=& F_{\mu\nu} e^{\mu}_{2}e^{\nu}_{0} \nonumber \\
      &=&  \frac{1}{rA^{1/2}}\left[(\Delta R)^{-1/2}\frac{1}{2}B_{0}
\left\{(\partial_{\theta}g_{t\phi})+\frac{2J}{M}\left(1 - \frac{\tilde{Q}}{2B_{0}J}\right)
(\partial_{\theta}g_{tt})\right\} \right.  \nonumber \\
&&\left. +(\Delta R)^{-1/2}\frac{2J}{r^3}\frac{1}{2}B_{0}
\left\{(\partial_{\theta}g_{\phi \phi})+\frac{2J}{M}\left(1 - \frac{\tilde{Q}}{2B_{0}J}\right)
(\partial_{\theta}g_{\phi t})\right\}\right], \nonumber \\
F_{30} &=& F_{\mu\nu} e^{\mu}_{3}e^{\nu}_{0} = 0, \\
F_{12} &=& F_{\mu\nu} e^{\mu}_{1}e^{\nu}_{2} = 0, \nonumber \\
F_{13} &=& F_{\mu\nu} e^{\mu}_{1}e^{\nu}_{3} \nonumber \\
      &=&  \left(\frac{\Delta}{S}\right)^{1/2} \frac{1}{rA^{1/2}\sin \theta}
\frac{1}{2}B_{0}\left\{(\partial_{r}g_{\phi \phi})+
\frac{2J}{M}\left(1 - \frac{\tilde{Q}}{2B_{0}J}\right)(\partial_{r}g_{\phi t})\right\}, \nonumber \\
F_{23} &=& F_{\mu\nu} e^{\mu}_{2}e^{\nu}_{3} \nonumber \\
      &=& \frac{1}{rA^{1/2}}\frac{1}{rA^{1/2}\sin \theta}
\frac{1}{2}B_{0}\left\{(\partial_{\theta}g_{\phi \phi})+
\frac{2J}{M}\left(1 - \frac{\tilde{Q}}{2B_{0}J}\right)(\partial_{\theta}g_{\phi t})\right\}
            \nonumber                          
\ea

\begin{center}
{\rm\bf V. The magnetic flux through the neutron star}
\end{center}

With the asymptotically uniform stationary axisymmetric magnetic field which is aligned
with the spin axis of a ``slightly charged'' slowly-rotating neutron star given above, we now would like
to compute the flux of the magnetic field across one half of the surface of the neutron star 
which is assumed to be of exact spherical geometry with radius $r_{0}$.
The physical motivation behind this study is to have some insight into the question
of how much of the magnetic flux can actually penetrate the surface of a rotating neutron star -
at least in idealized situations. We now begin by considering two vectors lying on the surface of
the neutron star which are given by
\be
dx_{1}^{\alpha}=(0, \; 0, \; d\theta, \; 0), \;\;\; dx_{2}^{\alpha}=(0, \; 0, \; 0, \; d\phi).
\ee
Then in terms of the 2nd rank tensor constructed from these two vectors [10],
\be
d\sigma^{\alpha\beta}=\frac{1}{2}(dx_{1}^{\alpha}dx_{2}^{\beta}-dx_{1}^{\beta}dx_{2}^{\alpha})
\ee
one now can define the invariant surface element of the neutron star as
\ba
ds &=& (2d\sigma_{\alpha\beta}\sigma^{\alpha\beta})^{1/2}\mid_{r_{0}} \\
   &=& (g_{\theta\theta}g_{\phi\phi})^{1/2}\mid_{r_{0}} \, d\theta d\phi. \nonumber 
\ea
Next, since the tensor $d\sigma^{\alpha\beta}$ is associated with the invariant surface
element of any (not necessarily closed) 2-surface, the flux of electric field and
magnetic field across any 2-surfaces can be given respectively by
\ba
\Phi_{E} &=& \int \tilde{F}_{\alpha\beta} d\sigma^{\alpha\beta} = \int \frac{1}{2}
  \epsilon_{\alpha\beta}^{\;\;\;\;\gamma\delta}F_{\gamma\delta}d\sigma^{\alpha\beta},
  \nonumber \\
\Phi_{B} &=& \int F_{\alpha\beta} d\sigma^{\alpha\beta}.  
\ea
In particular, the flux of magnetic field across one half of the surface of the neutron star is
\be
\Phi_{B} = \int_{r=r_{+}} F_{\alpha\beta} d\sigma^{\alpha\beta}
 = \int^{2\pi}_{0}d\phi \int^{\pi/2}_{0} d\theta F_{\theta\phi} \mid_{r_{0}}.
\ee
Then using $F_{\theta \phi}$ evaluated on the surface of the neutron star
\small
\ba
 F_{\theta\phi} \mid_{r_{0}} &=& \frac{1}{2}B_{0}\left[(\partial_{\theta}g_{\phi \phi})+
\frac{2J}{M}\left(1 - \frac{\tilde{Q}}{2B_{0}J}\right)(\partial_{\theta}g_{\phi t})\right]  \\
&=& \frac{1}{2}B_{0}\left\{2r^2_{0} - \frac{2J}{M}\left(1-\frac{\tilde{Q}}{2B_{0}J}\right)\frac{4J}{r_{0}}\right\}
\left[1 + \frac{J^2}{Mr^3_{0}}\left(1+\frac{2M}{r_{0}}\right) \right. \nonumber \\
&&\left. - \frac{5}{8} \frac{Q-J^2/M}{M^3}\left\{
+ \frac{2M}{[r^2_{0} (1-2M/r_{0})]^{1/2}}Q^{1}_{2}(\frac{r_{0}}{M}-1) - Q^{2}_{2}(\frac{r_{0}}{M}-1)\right\}\right]\sin \theta \cos \theta \nonumber \\
&+& \frac{3}{2}B_{0}\left\{2r^2_{0} - \frac{2J}{M}\left(1-\frac{\tilde{Q}}{2B_{0}J}\right)\frac{4J}{r_{0}}\right\}
\left[-\frac{J^2}{Mr^3_{0}}\left(1+\frac{2M}{r_{0}}\right) \right. \nonumber \\
&&\left. + \frac{5}{8} \frac{Q-J^2/M}{M^3}\left\{
+ \frac{2M}{[r^2_{0} (1-2M/r_{0})]^{1/2}}Q^{1}_{2}(\frac{r_{0}}{M}-1) - Q^{2}_{2}(\frac{r_{0}}{M}-1)\right\}\right](\sin \theta \cos^3 \theta - \sin^3 \theta \cos \theta ) \nonumber 
\ea
\normalsize
and the result of integration, $\int^{\pi/2}_{0}d\theta \sin \theta \cos \theta = 1/2$,
$\int^{\pi/2}_{0}d\theta (\sin \theta \cos^3 \theta - \sin^3 \theta \cos \theta) = 0$,
we finally arrive at
\ba
&&\Phi_{B} = \pi B_{0}\left\{r^2_{0} - \frac{4J^2}{Mr_{0}}\left(1-\frac{\tilde{Q}}{2B_{0}J}\right)\right\}\times  \\
&&\left[1 + \frac{J^2}{Mr^3_{0}}\left(1+\frac{2M}{r_{0}}\right) 
- \frac{5}{8} \frac{Q-J^2/M}{M^3}\left\{
+ \frac{2M}{[r^2_{0} (1-2M/r_{0})]^{1/2}}Q^{1}_{2}(\frac{r_{0}}{M}-1) - Q^{2}_{2}(\frac{r_{0}}{M}-1)\right\}\right] .
\nonumber
\ea
This is the magnetic flux through one half of the surface of the slowly-rotating neutron star with
accretion electric charge $\tilde{Q}$. Some discussions concerning the nature of this magnetic flux
are now in order :  \\

(i) In the absence of the accretion charge, i.e., $\tilde{Q}=0$, the magnetic flux above appears
to drop to zero (i.e., the magnetic field gets entirely expelled) when the neutron star is
maximally-rotating, which, based on the expression for the magnetic flux across the surface of
the neutron star given in eq.(39), amounts to $J^2_{max} = Mr^3_{0}/4$. This curious feature indeed
has its counterpart in the Kerr black hole case \cite{klk} which has been known for some time.
For the present case of neutron star, however, we are not fully qualified to make any definite
statement on this point since the Hartle-Thorne metric employed represents the region surrounding a
{\it slowly-rotating} neutron star. Thus the expression for the magnetic flux given above may be
invalidated for the case of rapidly-rotating neutron star and that is beyond the scope of the
present work as the associated metric is not available for now. \\

(ii) With the nonvanishing accretion charge having value in the range $0<\tilde{Q} <2B_{0}J$,
it is interesting to note that
\begin{eqnarray}
\Phi_{B}(\tilde{Q}\neq 0) > \Phi_{B}(\tilde{Q}= 0).
\end{eqnarray}
Actually, this is one of the points of central importance we would
like to make in the present work. The physical interpretation of this characteristic
can be briefly stated as follows. Once the neutron star accretes charge, toroidal currents occur
as a result of rotation of the star. This toroidal currents, in turn, generates new magnetic  
fields which would be additive to the existing ones increasing the total magnetic flux through
the surface of the rotating neutron star. As has been studied in detail in \cite{hongsu} recently,
essentially the same phenomenon takes place in the case of slightly charged Kerr black hole.
There, the interpretation was even clearer ; when the spin of the hole and the asymtotically
uniform magnetic field are parallel (antiparallel), the hole selectively accretes
positive (negative) charges according to the injection energy argument proposed by Carter and 
they, in turn, generate magnetic fields which are clearly additive to the existing ones. 
And we expect that essentially the same will be true for the present rotating neutron star case
although we could not determine the sign of accretion charge in association with the relative
direction between the spin of the neutron star and the direction of the magnetic field due to
the highly complex structure of Hartle-Thorne metric as we mentioned earlier. \\

(iii) For the Wald charge, $\tilde{Q}=2B_{0}J$, the magnetic flux across the surface of the
rotating neutron star appears to take on rather special value
\ba
&&\Phi_{B} = B_{0}\pi r^2_{0}\times  \\
&&\left[1 + \frac{J^2}{Mr^3_{0}}\left(1+\frac{2M}{r_{0}}\right) 
- \frac{5}{8} \frac{Q-J^2/M}{M^3}\left\{
+ \frac{2M}{[r^2_{0} (1-2M/r_{0})]^{1/2}}Q^{1}_{2}(\frac{r_{0}}{M}-1) - Q^{2}_{2}(\frac{r_{0}}{M}-1)\right\}\right] .
\nonumber
\ea
As mentioned in the introduction, it may be interesting to compare this analysis of the nature of
the magnetic flux through a rotating neutron star with that of the Kerr hole studied in \cite{hongsu}.

\begin{center}
{\rm\bf VI. Concluding remarks}
\end{center}
  
In the present work, based on the solution-generating method given by Wald, the magnetic fields around
both the uncharged and (slightly) charged neutron star have been obtained. Particularly for the
charged neutron star, it has been demonstrated following again the argument by Wald that 
the neutron star will gradually accrete the charge until it reaches the equilibrium value 
$\tilde{Q}=2B_{0}J$. In association with the ``magnetic braking'' model of Goldreich and Julian \cite{gj}
for the spin-down of magnetized rotating neutron stars, the magnetic flux through one half of the
surface of the rotating neutron star has been computed as well. With the nonvanishing accretion 
charge having value in the range $0 < \tilde{Q} \leq 2B_{0}J$, the total magnetic flux through the 
neutron star has been shown to be greater than that without the accretion charge.
Indeed, the physical interpretation of this characteristic
can be briefly stated as follows. When the spin of the neutron star and the direction of the asymtotically
uniform magnetic field are parallel (antiparallel), the neutron star would selectively accrete
positive (negative) charges following the injection energy argument proposed by Carter and 
they, in turn, generate magnetic fields additive to the existing ones.   \\
Next, one might be worried about the validity of the Hartle-Thorne metric for the region 
surrounding the slowly-rotating neutron stars employed in this work to describe the magnetosphere 
of pulsars which seem rapidly-rotating having typically millisecond pulsation periods.
Here the ``slowly-rotating'' means that the neutron star rotates relatively slowly compared to
the equal mass Kerr black hole which can rotate arbitrarily rapidly up to the maximal rotation
$J = M^2$. Thus this does not necessarily mean that the Hartle-Thorne metric for slowly-rotating
neutron stars cannot properly describe the millisecond pulsars. To see this, note that according
to the Hartle-Thorne metric, the angular speed of a rotating neutron star is given by the 
Lense-Thirring precession angular velocity in eq.(9) at the surface of the neutron star, which,
restoring the fundamental constants to get back to the gaussian unit, is 
\ba
\omega = \frac{2J}{r^3_{0}}\left(\frac{G}{c^2}\right), ~~~{\rm with}
~~J = \tilde{a}M^2\left(\frac{G}{c}\right) ~~(0<\tilde{a} <1).
\ea
Thus the Hartle-Thorne metric gives the angular speed of a rotating neutron star, having the 
data of a typical radio pulsar, $M\sim 2\times 10^{33}(g)$, $r_{0}\sim 10^6 (cm)$, as
$\omega = 2\tilde{a}(M^2/r^3_{0})(G/c)(G/c^2) \sim 10^3 (1/sec)$ which, in turn, yields the
rotation period of $\tau = 2\pi/\omega \sim 10^{-2} (sec)$. And here we used,
$(G/c^2) = 0.7425\times 10^{-28} (cm/g)$ and $(G/c) = 2.226\times 10^{-18} (cm^2/g\cdot sec)$.
Indeed, this is impressively
comparable to the observed pulsation periods of radio pulsars $\tau \sim 10^{-3} - 1 (sec)$
we discussed earlier. As a result, we expect that the Hartle-Thorne metric is well-qualified to
describe the geometries of millisecond pulsars. \\
Lastly, we mention an inherent drawback of our treatment of the issues presented in this work.
In the present work, we have mainly considered the case with symmetric geometry in which the 
stationary axisymmetric magnetic field is precisely aligned with the neutron star's axis of rotation.
Although this simplification/idealization was inevitable in order to make the system mathematically
tractable, this would render the magnetized rotating neutron star fail to describe pulsars
in the real world as the pulsar involves, by definition, oblique spin-magnetic field configuration.
(In the case involving a rotating black hole, however, such oblique configuration cannot be
maintained since any possible initial misalignment between the spin and the direction of the
magnetic field will be removed in an efficient manner either by a viscous torque when the accretion
disc is involved \cite{bp} or by a magnetic alignment torque \cite{align}.)
Indeed, almost all of the theoretical studies of pulsar electrodynamics assume this idealized
symmetric geometry for concrete, analytical calculations and concerning this limitation,
our work makes no progress in this direction. Thus we are still away from the satisfactory theoretical description
of the real physics that takes place in the pulsar magnetosphere. Nevertheless
it is our hope that the result of the present work, obtained in a more realistic situation when 
some amount of plasma (charges) are around in addition to the magnetic field, 
eventually lends supports to the operational nature of the Goldreich-Julian mechanism of the
magnetic braking to explain the pulsar radiation due to its spin-down and puts it on a firmer ground at least on the 
theoretical side.

\vspace*{1cm}

\begin{center}
{\rm\bf Acknowledgements}
\end{center}

H.Kim was financially supported by the BK21 Project of the Korean Government. 
CHL and HKL were supported in part by grant No. R01-1999-00020 from the Korea
Science and Engineering Foundation.

\end{document}